\documentclass[aps,prb,reprint,superscriptaddress,showpacs,floatfix,twocolumn]{revtex4-1}
\usepackage{graphicx}% Include figure files
\usepackage{dcolumn}% Align table columns on decimal point
\usepackage{bm}% bold math
\begin{document}

%\preprint{}

\title{Topological fate of edge states of Bi single bilayer on Bi(111)}


\author{Han Woong Yeom}
\email{yeom@postech.ac.kr}
\affiliation{Center for Artificial Low Dimensional Electronic Systems, Institute for Basic Science (IBS), 77 Cheongam-Ro, Pohang 790-784, Republic of Korea}
\affiliation{Department of Physics, Pohang University of Science and Technology (POSTECH), Pohang 790-784, Republic of Korea}

\author{Kyung-Hwan Jin}
\affiliation{Department of Physics, Pohang University of Science and Technology, Pohang 790-784, Korea}

\author{Seung-Hoon Jhi}
\affiliation{Department of Physics, Pohang University of Science and Technology, Pohang 790-784, Korea}


\date{\today}

\begin{abstract}
We address the topological nature of electronic states of step edges of Bi(111) films by first principles band structure calculations. We confirm that the dispersion of step edge states  reflects the topological nature of underlying films. This result unambiguously denies recent claims that the step edge state on the surface of a bulk Bi(111) crystal or a sufficiently thick Bi(111) films represents non-trivial edge states of the two dimensional topologcial insulator phase expected for a very thin Bi(111) film. The trivial step edge states have a gigantic spin splitting of one dimensional Rashba bands and the substantial intermixing with electronic states of the bulk, which might be exploited further.
\end{abstract}

% insert suggested PACES numbers in braces on next line

%\keywords{}

\maketitle


\section{INTRODUCTION}
    The hallmark of a topological insulator (TI) is the existence of chiral edge mode(s) along edges of the material. In its two dimensional (2D) version, the chiral edge mode is highly interesting, representing a quantum spin Hall (QSH) channel.\cite{Bernevig,Konig,Liu,Knez} Very recently, such QSH edge states were reported to be directly probed by scanning tunneling microscopy and spectroscopy (STM/STS) for Bi single bilayers (BL) grown on Bi$_2$Te$_2$Se\cite{Kim,Sabater} and graphene,\cite{Lu} following an earlier theoretical prediction.\cite{Murakami,Wada} The QSH phase of Bi ultrathin films was predicted to be stable up to 8 BL thickness\cite{Liu2} and a thicker film becomes topologically trivial as expected for bulk Bi crystals.\cite{Takayama} On the other hand, a recent STS study found edge-localized states, possibly spin-polarized, along step edges at the surface of a Bi(111) crystal and attributed them as the topological edge state of the 2D topologically insulating (or semimetal) Bi single BL.\cite{Yazdani} This claim was followed by a more recent work on a thick Bi(111) films grown on a Si substrate.\cite{Kawakami} That is, these works interpret the surface layer of a Bi(111) crystal as a 2D topological material in apparent contradiction to the current knowledge of a Bi crystal as a topologically trivial metal. This leads to important questions of whether the topologically trivial Bi(111) crystal is covered with a non-trivial 2D layer and how the interaction of the surface Bi layer with its bulk preserves (or affects) its topological property. 
    
In this work, we explicitly check the 1D band structure of step edges of Bi(111) films as a function of the film thickness. We unambiguously clarify that the Bi(111) surface layer on top of a sufficiently thick film becomes a trivial 2D semimetal, reflecting the topological phase transition of the Bi(111) film, and its step edge has trivial spin-split bands of the 1D Rashba type. 
%------------------------------------------ Figure 1 ------------------------------------------
\begin{figure}
\includegraphics{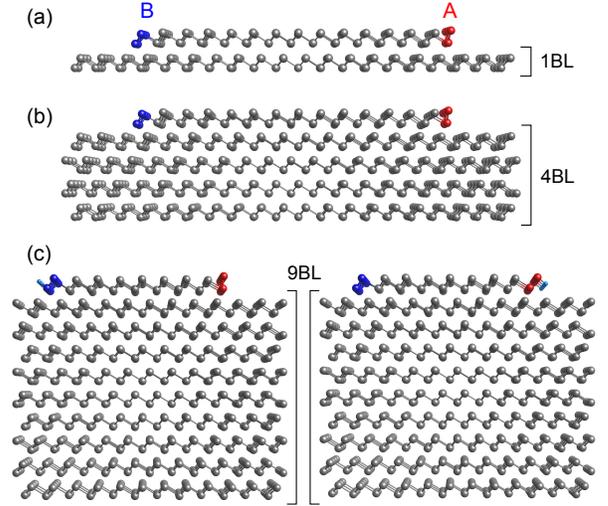}
\caption{\label{fig:epsart}(color online) Structural models of zigzag Bi nanoribbons on (a) 1, (b) 4, and (c) 9 BL Bi(111) films. Type A and type B edges are marked by red and blue atoms respectively, which undertake different types of reconstructions. In order to reduce the computational load, we used narrower nanoribbons on 9 BL films. In this case, one edge is passivated by H atoms to ensure the decoupling between.}\label{fig1}
\end{figure}
%----------------------------------------------------------------------------------------------

\section{CALCULATIONS}
\textit {Ab initio} calculations were carried out in the plane-wave basis within the generalized gradient approximation for the exchange-correlation functional \cite{Kresse, Perdew}. A cut-off energy of 400~eV was used for the plane-wave expansion and a \textit{k}-point mesh of 15$\times$1$\times$1 for the Brillouin zone sampling. In order to investigate edge states, we carried out calculations for a single Bi nanoribbons (BNR) of 9 or 15 Bi zigzag chains on top of 1$\sim$6 or 7$\sim$9 BL slabs, respectively, as illustrated in Fig. 1. This mimics step edges with the zigzag structure of 2-10 BL films. Atoms of Bi slabs and nanoribbons on top were fully relaxed until the Helmann-Feynman force was less than 0.01~eV/\AA. 
%------------------------------------------ Figure 2 ------------------------------------------
\begin{figure*}
\includegraphics{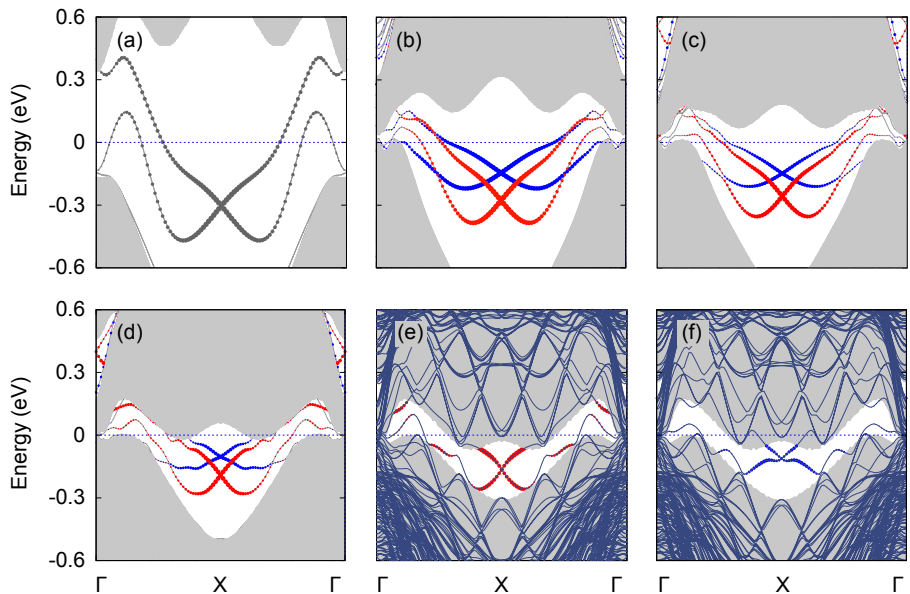}
\caption{(color online) Calculated band structure (a) pristine floating 1 BL Bi nanoribbon and 1 BL Bi nanoribbons on top of (b) 1, (c) 2, (d) 4, (e)-(f) 9 BL films. Edge states are distinguished by red and blue dots for the A and B type edges according to Fig. 1. Shaded regions are the projection of the bulk (whole film) bands of 1, 3, 5 and 10 BL films, respectively.}\label{fig2}
\end{figure*}
%----------------------------------------------------------------------------------------------

\section{RESULTS and DISCUSSION}
For floating Bi films, detailed band structure calculations are already available, which tell that the 2D TI nature of Bi bilayers (BL) is maintained only up to 8 BL.\cite{Liu2} Thicker films become metallic with the band gap closed, obtaining the bulk property of a topologically trivial metal. The closure of the band gap in this case does not involve the band inversion, which is another hallmark of TI. We reproduce this result consistently with our own calculations. We also examine the band structure of surface layers of these films. Surface layers of a thick Bi(111) film or the bulk Bi(111) have topologically trivial 2D spin helical bands due to the Rashba-type spin-orbit interaction.\cite{Koroteev,Koroteev2} In contrast to thin floating Bi bilayers, the bands of a Bi(111) surface layer have no band gap and no band inversion near the Fermi energy. Since the band inversion is required to have a topologically non-trivial edge state, there is no obvious reason \textit{a priori} to expect any topological edge state on the edge of these surface layers. The origin of the difference between the floating Bi bilayer and the Bi(111) surface layer will be discussed further below. 

We then calculated the edge band structure explicitly by simulating the step edge structure of the surface layer on Bi(111) films of varying thickness. As shown in Fig. 1, we generate nanoribbons of 9-15 unitcell wide on top of Bi(111) films up to 9 BL thickness. The thickness of 9 BL is confirmed to be sufficient to simulate the topologically trivial bulk property. Since the whole atoms are freely relaxed, the edge atoms are substantially reconstructed to reduce the energy cost of the dangling bond creation. Two different edge configurations and reconstructions are formed due to the BL structure; edge atoms are buckled down or up for so called A or B type edge, respectively, as shown in Fig. 1. These two edge structures yield slightly different edge state dispersions as shown in Fig. 2. Commonly for these two edges, the edge bands have Dirac-like crossings at the edge of Brillouin zone (X) within the enlarged band gap. These crossed bands have a helical spin texture (as confirmed in our spin-polarized calculations, data not shown here). This spin helicity and the Dirac-like crossings are indeed very similar to the QSH edge state of the 2D TI phase of a single BL film [Fig. 2(a)].
%------------------------------------------ Figure 3 ------------------------------------------
\begin{figure}
\includegraphics{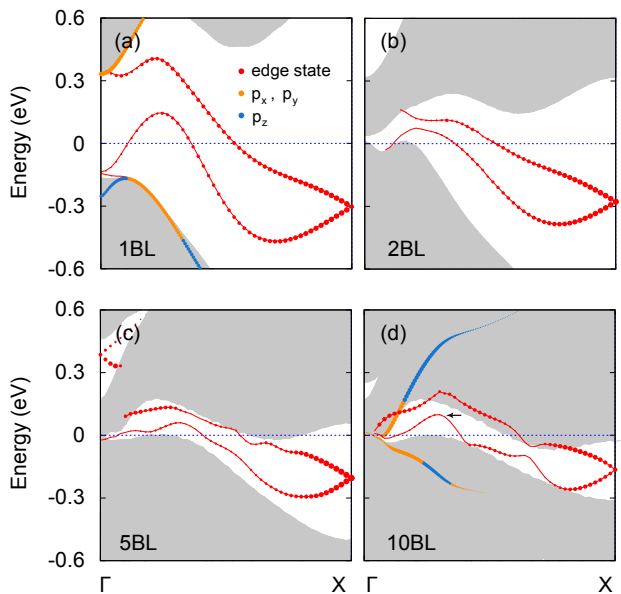}
\caption{(color online) Band structures of Bi nanoribbons on various Bi(111) thin films. (a) Edge states of a zigzag-edged floating Bi nanoribbon. (b)-(d) Edge states of the type B edges (the same type of edge as in Ref. 12) on 1, 4 and 9 BL films. Shaded regions are the projection of the corresponding bulk (whole film) bands. For the floating or surface layer of 1 and 10 BL films, orbital characters of the valence and conduction band edges are indicated.}\label{fig3}
\end{figure}
%----------------------------------------------------------------------------------------------

However, an important change can be noticed even for the 2 BL case, the nanoribbon on a single BL film substrate [Fig. 2(b)]. Even at this thickness, the band gap of the film reduces substantially, especially at the center of the Brillouin zone (${\Gamma}$), which closes down completely above 8 BL substrate. An immediate consequence of the band gap closure is that the topological nature of the step edge state becomes apparently ambiguous even on the 1 BL substrate; as the band gap closes, the upper and lower branches of the spin split edge states get very close to each other at ${\Gamma}$ while their Dirac-like crossing at X is preserved. If these two branches merges at ${\Gamma}$, then their dispersions become topologically trivial and the Dirac-like crossings at X becomes trivial Rashba band crossings. In fact, this change is what is expected from the topological phase transition of the film at a thickness around 8 BL. For a thicker substrate than 4 BL, the band gap narrowing is substantial in the whole Brillouin zone and the edge states hybridize with the substrate bands for a large part of the Brillouin zone. Thus, the step edge state is not truly localized along the edge atoms except for a limited part of the Brillouin zone. Note that these changes do not depend on the type of the edge structures. 

As mentioned above, the crucial part for the topological nature of the edge state dispersion is the band dispersion near ${\Gamma}$. We thus scrutinize the change of edge state dispersions as shown in Fig. 3 for one type, type B, of the edge structure. As clearly shown here, the step edge states on a sufficiently thick film have their spin-split bands merged at the ${\Gamma}$ point, becoming a trivial Rashba spin-split bands. The same conclusion is reached for the other type of edges. We confirm that this change occurs concomitantly with the quantum phase transition from the 2D topological insulator to a 3D trivial semimetal. This can be shown by the disappearance of the band inversion between the bands of different \textit{p} orbitals; as depicted in Figs. 3(a) and 3(d), the valence and conduction bands at ${\Gamma}$ have the clear band inversion for the single floating BL film but return back to normal on the 9 BL substrate.
%------------------------------------------ Figure 3 ------------------------------------------
\begin{figure}
\includegraphics{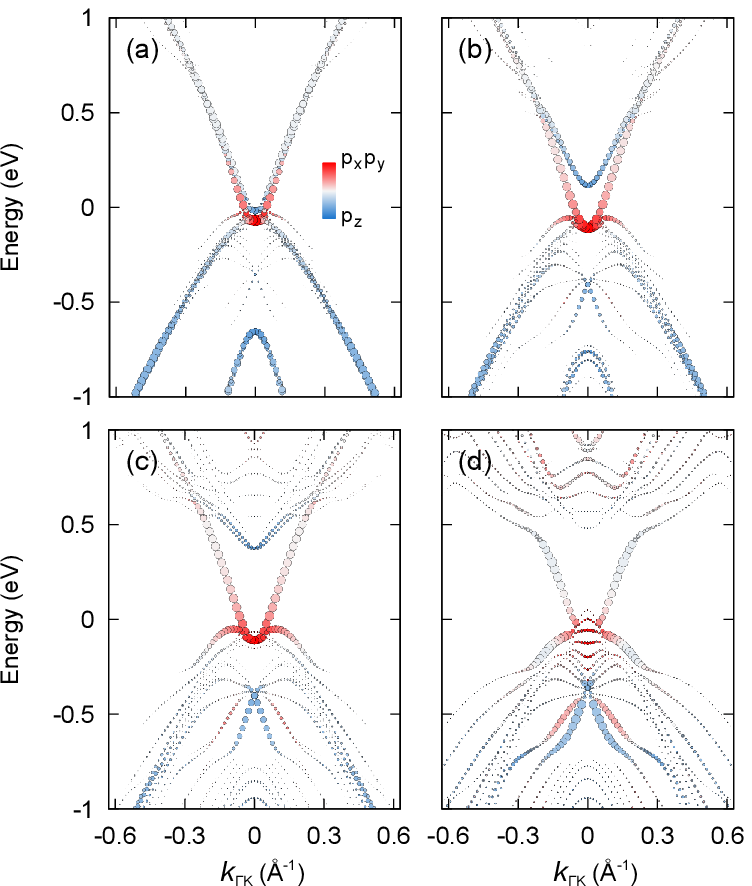}
\caption{(color online) (a)-(d) Calculated band structures of a sing Bi bilayer on top of 9 BL films at a distance of 4, 3, 1, and 0 $\AA$ from the equilibrium distance, respectively. The states originating from the Bi bilayer are marked by red (\textit{px} and \textit{py} orbitals) and blue (\textit{pz} orbital). The in-plane lattice constant is fixed with that of the equilibrium surface layer case.}\label{fig4}
\end{figure}
%----------------------------------------------------------------------------------------------

In contrast to the present work, the previous STS measurements claimed the topologically non trivial step edge state on top of a bulk Bi(111) crystal.\cite{Yazdani} What was actually measured is a part of the dispersion of the edge localized state, which is indicated by an arrow in Fig. 3(d). They claimed the spin polarized nature of this part of the band, which is obviously true due to the Rashba spin splitting. The topological nature of the edge state is, however, determined not by the spin splitting but by their dispersions especially at ${\Gamma}$ in the present case. While one can agree that this Rashba-type edge states evolve into QSH edge states at a very small thickness as already observed on single BL films on Bi$_2$Te$_2$Se and graphene,\cite{Kim,Sabater,Lu} the bands on a thick film are obviously not topological edge states but 1D bands with the Rashba spin splitting. The 1D chiral electron state itself with a large Rashba splitting was previously introduced in a surface nanowire array.\cite{Park} In other words, the strong coupling of the electronic states of a single BL surface layer with those of its substrate destroys their topological nature. This substrate effect is detailed further in Fig. 4. We calculate the band dispersions of a Bi single BL at a varying distance from its equilibrium position as a surface layer of 9 BL film. Note that, even at a sufficiently large distance from the bulk with a minimal interaction, the single BL film loses the insulating property of a free standing film. This is due to the substantial difference in the lattice constant between the free standing film and the bulk (about 5 $\%$ expansion in the bulk). Even with the closed band gap, the non-interacting film maintains its topological property with the band inversion near the Fermi level as in the free standing film. When the interaction with the bulk is turned on at a shorter distance, the band inversion disappears with bands of a single character (red colored bands) across the Fermi level and the valence bands of the non-interacting film (blue bands) hybridized strongly with those of the bulk. These results show clearly how the single BL film and the Bi(111) substrate interact. The existence of a spin-helical edge state is necessary for a TI but not a sufficient condition to dictate the TI property. Unfortunately, this simple point was ignored by the previous STS work as well as the substantial change of the electronic state of the Bi bilayer on the bulk substrate.\cite{Yazdani}

\section{CONCLUSION}
We investigate explicitly the topological nature of electronic states of step edges of Bi(111) films by first principles calculations. We show that the dispersion of step edge states reflects the topological nature of underlying films, unambiguously denying the existence of topological edge states along step edges on surfaces of a bulk Bi(111) crystal or a sufficiently thick Bi(111) film. The trivial step-edge states have spin splitting of one dimensional Rashba bands. This 1D band with a huge Rashba splitting might be interesting for the search of Majorana Fermion but the existence of a strong intermixing with the substrate electronic state has to be considered.

\begin{acknowledgments}
This work was supported by Institute for Basic Science (Grant No. IBS-R015-D1). 
\end{acknowledgments}


\end{document}